\documentclass[aps,prl, twocolumn,showpacs]{revtex4}

\usepackage{amsmath}
\usepackage{graphicx}
\usepackage{times}
\usepackage{amssymb}
\usepackage{hyperref}
\usepackage{textcomp}
\usepackage{color}

\begin{document}
\title{Entanglement in atomic resonance fluorescence}
\author{P. Gr\"unwald\footnote{Electronic address: peter.gruenwald2@uni-rostock.de} and W. Vogel\footnote{Electronic address: werner.vogel@uni-rostock.de}} \affiliation{Arbeitsgruppe Quantenoptik, Institut f\"ur Physik, Universit\"at Rostock, 18055 Rostock, Germany}
\date{received \today}

\begin{abstract}
The resonance fluorescence from regular atomic systems is shown to represent a continuous source of non-Gaussian entangled radiation propagating in two different directions. For a single atom entanglement occurs under the same conditions as squeezing. For more atoms, the entanglement can be more robust against dephasing than squeezing, hence providing a useful continuous  source for various applications of entangled radiation. 
\end{abstract}


\maketitle

The resonance fluorescence of a single atom played an outstanding 
role when searching for a radiation source that clearly displays the quantum nature of light. A single two-level atom, after being excited by a driving laser, can only emit a single photon at once and then it must be re-excited, before another photon can be emitted. This feature led to the prediction of the nonclassical effect of photon antibunching~\cite{carm}, which could first be demonstrated in the resonance fluorescence of an atomic beam~\cite{ki-dag-ma} and later with a trapped ion~\cite{walther}.

Sub-Poissonian photon statistics could also be observed for the first time in resonance fluorescence~\cite{short} and squeezing in resonance fluorescence was predicted~\cite{walls}. Even the resonance fluorescence of many atoms 
can show squeezing, which requires stable phase relations between the emitters. This can be achieved by a regular arrangement of the atoms~\cite{vowe}, or by detection of the fluorescence in the forward direction with respect to the pump beam~\cite{heire}. 
Squeezing from strongly driven regular atomic systems has also been studied~\cite{Keitel}.
In an early experiment with regular atoms, the 
interference of the fluorescence of two trapped ions was demonstrated~\cite{wineland}. Squeezing in resonance fluorescence could be observed for samples of many atoms~\cite{LuBali}. 

A direct demonstration of squeezing in the resonance fluorescence of a single atom has not been realized yet. For this purpose it was proposed to apply homodyne correlation measurements~\cite{vo}. The method has been further developed to detect general field correlation functions by balanced homodyne correlation techniques~\cite{sh-vo}. This opens possibilities to study the most general nonclassical features of the atomic resonance fluorescence radiation~\cite{vo-08}. As an example, intensity-field correlation functions could be measured in resonance fluorescence~\cite{blatt}, which is a first step in such a direction.  

In the context of applications for quantum information processing, among the manyfold of nonclassical effects entanglement became of particular importance, for recent reviews see~\cite{horodecki,guehne}.
A variety of possibilities to create entanglement in atoms, e.g. via cooperative fluorescence, has been studied, see~\cite{FicekRep} and references therein. A cold atom in a cavity can serve as a stable source for entangled EPR-type photons, by utilizing the coupling between the scattered photons and the quantized atomic center-of-mass motion~\cite{vitali}.
To our best knowledge, however, the resonance fluorescence radiation itself has not been studied yet with respect to entanglement.

In the present Letter we study the resonance fluorescence of a regular system of atoms, separated from each other by several wavelengths of the driving laser. Entanglement is found in the electric field strengths of the fluorescence radiation propagating in different directions from the atomic sample. Remarkably, the robustness of the entanglement against dephasing of the atomic transitions increases with increasing number of atoms. This 
renders it possible to design continuously radiating sources of non-Gaussian  entangled light for various applications.

For simplicity, we deal with
the resonance fluorescence of atomic two-level systems. The theoretical background is given in several textbooks, e.g.~\cite{Vogel:06}. Let us consider an ensemble of $N$ non-interacting two-level atoms, 
located at the positions $\mathbf r^{(n)}$ $(n=1\ldots N)$.
The electric field strength operator at position $\mathbf r$ and time $t$ is given in source-quantity representation as
\begin{equation}
 \hat{\mathbf E}^{(\pm)}(\mathbf r,t)=\hat{\mathbf E}^{(\pm)}_{\rm{f}}(\mathbf r,t){+}\sum_{n=1}^N\hat{\mathbf E}^{(\pm)}_{{\rm s},n}(\mathbf r,t), \label{eq.E-field}
\end{equation}
where the indices 'f' and 's' denote the free and source field parts, respectively. The superscript $(\pm)$ indicates positive/negative frequency field components. 
For our purposes, the geometry is chosen such that the pump laser does not hit the detectors, hence the corresponding free-field contribution is in the vacuum state, $\langle\cdots\hat{\mathbf E}^{(+)}_{\rm{f}}\rangle{=}\langle\hat{\mathbf E}^{(-)}_{\rm{f}}\cdots\rangle{=}0$. In the normally ordered correlation functions considered in the following, we can throughout omit the free field contributions in Eq.~(\ref{eq.E-field}) and only keep track of the source fields. 

The source field part of the resonance fluorescence field of a two-level atom in rotating wave approximation is given by
\begin{equation}
        \hat{\mathbf E}^{(+)}_{{\rm s},n}(\mathbf r,t)=\mathbf g(\mathbf r-\mathbf r^{(n)}) \hat A_{12}^{(n)}(t^{(n)}),
\end{equation}
with $\hat A_{ab}^{(n)}= |a^{(n)}\rangle\langle b^{(n)}|$ $(\{a,b\}{=}1,2)$ being the atomic flip operator of the $n$-th atom at the retarded time $t^{(n)}=t{-}|\mathbf r{-}\mathbf r^{(n)}|/c$. The function $\mathbf g(\mathbf r{-}\mathbf r^{(n)})$ relates the atomic source operators to the source field parts of the radiation field. Assuming the scattering center of the atoms at $\mathbf r =0$ and applying the far field approximation, $\mathbf g$ reads as
\begin{equation}
\nu \ \mathbf g(\mathbf r)=\frac{\mathbf d}{|\mathbf d|}-\frac{(\mathbf d\cdot\mathbf r)\mathbf r}{|\mathbf d||\mathbf r|^2}\label{eq.g},
\end{equation}
where $\mathbf d$ is the atomic dipole moment and  $\nu$ a scaling factor.

The atomic flip operators $\hat A_{ab}^{(n)}$ obey the equal time relation $\hat A_{ab}^{(n)}\hat A_{cd}^{(n)}{=}\delta_{bc}\hat A_{ad}^{(n)}$, with $\delta$ being the Kronecker-$\delta$. Their quantum averages are obtained  from the optical Bloch equations for the density matrix of a single atom. One has to include phase factors $\exp(-i\varphi_j^{(n)})$, according to~\cite{vowe}, to account for the interference between all atoms and the positions of the detectors. Here $n$ ($n=1,\dots, N$) numbers the atoms and $j$ labels the positions ${\bf r}_j$ of the used points of observation.
Since the atoms are separated by linear distances large compared to the pump-laser wavelength, the correlation functions of products of atomic operators of different atoms factorize, $\langle\hat A_{ab}^{(n)}\hat A_{a'b'}^{(m)}\rangle{=}\langle\hat A_{ab}^{(n)}\rangle\langle\hat A_{a'b'}^{(m)}\rangle$ for $m{\neq} n$. That is, under such conditions cooperative effects are negligible.

In the balanced homodyne correlation measurements under consideration, the interference of the signal field with the local oscillator is the observed signal, for details see~\cite{sh-vo}. The recorded correlation functions contain the slowly-varying signal field operators, 
$\hat{\tilde{\mathbf E}}^{(\pm)}(\mathbf r,t)=\hat{\mathbf E}^{(\pm)}(\mathbf r,t) e^{\pm i\omega_{\rm lo} t}$, where $\omega_{\rm lo}$ is the frequency of the local oscillator. The fields at the points of observation $\mathbf r_j$ (in the following we choose $j=1,2$) can be 
related to the Bosonic operators $\hat{{a}}_{j,\lambda}$ as 
\begin{equation}
\label{eq:field}
\hat{\mathbf {E}}^{(+)}(\mathbf r_j ,t) =\sum_\lambda \mathbf f_\lambda(\mathbf r_j) \hat{{a}}_{j,\lambda}(t). 
\end{equation}
The mode functions $\mathbf f_\lambda(\mathbf r_{j})$ describe those fields which are recorded by homodyne correlation measurements within the spatio-temporal resolution of the detectors.

An important class of entangled states can be characterized by the negativity of the partial transposition of the quantum state under study~\cite{Peres}. The Peres condition for entanglement has been reformulated as~\cite{sh-vo-m}
\begin{equation}
\label{eq:peres-sv}
\langle (\hat{f}^\dagger \hat{f})^{\rm PT}\rangle<0, 
\end{equation}
in terms of the partial transposition of an operator $\hat{f}^\dagger \hat{f}$, with $\hat{f}$ being a function of the Bosonic operators describing the system~\cite{sh-vo-m}, for the multi-partite extension see~\cite{sh-vo-mm}. Expanding the operator $\hat{f}$ into a Taylor series of $\hat a_{j}$, the entanglement conditions lead to a hierarchy of negativity conditions in terms of minors whose entries are moments of $\hat a_{j}$ and $\hat a_{j}^{\dagger}$.
For a bipartite system consisting of two modes, a simple example of such an entanglement condition is
\begin{equation}
%D^{\mathbf N_0}=
\left|
\begin{array}{ccc}
1 &\langle\hat a_1\rangle &\langle\hat a_2\rangle\\
\langle\hat a_1^{\dagger}\rangle & \langle\hat a_1^{\dagger}\hat a_1\rangle &\langle \hat a_1^{\dagger}\hat a_2\rangle\\
\langle a_2^{\dagger}\rangle & \langle \hat a_2^{\dagger}\hat a_1\rangle &\langle \hat a_2^{\dagger}\hat a_2\rangle
\end{array}\right|<0. \label{eq.Dk0}
\end{equation}
This condition can be generalized for the multi-mode fields defined in Eq.~(\ref{eq:field}). 
Now the operator $\hat{f}$ in the condition~(\ref{eq:peres-sv}) is chosen as a Taylor series in terms of the fields in two space-time points $(i=1,2)$, 
$\hat{\mathbf {E}}^{(\pm)}(\mathbf r_j ,t)$, for which we use
the short hand $\hat{\mathbf {E}}^{(\pm)}_j$. 
We can again formulate the entanglement conditions in terms of a hierarchy of minors. The quantum-field theoretical counterpart of the two-mode  entanglement condition~(\ref{eq.Dk0}) reads, in terms of the minor $\mu$, as
\begin{equation}
\mu \equiv
\left|
\begin{array}{ccc}
1 &\langle \hat{\mathbf {E}}^{(+)}_1\rangle &\langle \hat{\mathbf {E}}^{(+)}_2\rangle\\
\langle \hat{\mathbf {E}}^{(-)}_1\rangle & \langle \hat{\mathbf {E}}^{(-)}_1 \hat{\mathbf {E}}^{(+)}_1\rangle &\langle \hat{\mathbf {E}}^{(-)}_1 
\hat{\mathbf {E}}^{(+)}_2\rangle\\
\langle \hat{\mathbf {E}}^{(-)}_2\rangle & \langle \hat{\mathbf {E}}^{(-)}_2\hat{\mathbf {E}}^{(+)}_1\rangle &\langle \hat{\mathbf {E}}^{(-)}_2
\hat{\mathbf {E}}^{(+)}_2\rangle
\end{array}\right|<0. \label{eq.Dk1}
\end{equation}
The occurring correlations of the fields in two space-time points relative to the radiating source describe the measured behavior for the often used assumption of the detector areas being small compared to the spatial coherence area of the fields and the time resolution being smaller than the coherence time. For practical applications the corresponding fields can also be collimated by apt optics and propagated to two different regions, e.g. to Alice and Bob, which are far apart from each other and far from the source.
Both these possibilities, the detection and the propagation of the entangled fields, are specified by choosing the mode functions in Eq.~(\ref{eq:field}) to describe localized and propagating fields, respectively.

Expanding the minor $\mu$ in Eq.~(\ref{eq.Dk1}), one readily gets as condition for entanglement
\begin{equation}
\label{eq:ent-cond}
\langle\Delta\hat{\mathbf {E}}_1^{(-)}\Delta\hat{\mathbf {E}}_1^{(+)}\rangle\langle\Delta\hat{\mathbf {E}}_2^{(-)}\Delta\hat{\mathbf {E}}_2^{(+)}\rangle{-}|\langle\Delta\hat{\mathbf {E}}_1^{(-)}\Delta\hat{\mathbf {E}}_2^{(+)}\rangle|^2 <0.
\end{equation}
Here we have introduced the symbol $\Delta \hat O=\hat O-\langle\hat O\rangle$ for an arbitrary operator $\hat O$. 
If one would further reduce the minor in the condition~(\ref{eq.Dk1}) to  
the dimension 2x2, it is easy to see that the resulting minors  are always non-negative. Hence the entanglement condition~(\ref{eq:ent-cond}) under study is a rather simple example, but it will turn out in the following to be sufficient for demonstrating entanglement in resonance fluorescence.

For the resonance fluorescence from identical two-level atoms, the dependence on the number $N$ of atoms in $\mu$ is in the phase factors $\varphi_j^{(n)}$. 
After some straightforward algebra, the left hand side of the condition~(\ref{eq:ent-cond}) reads as 
\begin{align}
        &\mu 
=\mathbf g_1^2\mathbf g_2^2\Bigg[N^2\left(\sigma_{22}-|\sigma_{21}|^2\right)^2\nonumber\\
        &{-}\cos^2(\vartheta)\left(\sigma_{22}-|\sigma_{21}|^2\right)^2\sum_{m,n=1}^N e^{i(\varphi_1^{(n)}-\varphi_2^{(n)}-\varphi_1^{(m)}+\varphi_2^{(m)})}\nonumber\\     &-\sin^2(\vartheta) |\sigma_{21}|^4\sum_{m,n,k,l=1}^Ne^{i(\varphi_1^{(n)}
-\varphi_2^{(m)}-\varphi_1^{(k)}+\varphi_2^{(l)})}\Bigg],\label{eq.statD1}
\end{align}
where $\sigma_{ba}{=}\langle\hat A_{ab}\rangle$ are the slowly varying density matrix elements of the identical two-level atoms and $\vartheta$ is the angle between the two functions $\mathbf g_1$ and $\mathbf g_2$, with
$\mathbf g_j=\mathbf g(\mathbf r_j)$.
For random positions of the atoms, the phase combinations in the second and third line in Eq.~(\ref{eq.statD1}) are averaging to zero, so that only the first line contributes. Under these conditions the minor $\mu$ is always positive semidefinite, and hence entanglement does not exist or it cannot be identified by the condition~(\ref{eq:ent-cond}).

Let us consider now the situation for a regularly arranged sample of $N$ atoms, whose resonance fluorescence is known to show squeezing~\cite{vowe},
contrary to the situation for randomly distributed atoms.
Under optimal conditions all the phase differences, $\varphi_{i}^{(n)}{-}\varphi_{j}^{(m)}$, vanish. This requires that all atoms lie on the central plane between the two detector positions, and the (not necessarily equal) distances between the atoms are  integer multiples of the pump laser wavelength. In such an optimal configuration, the condition~(\ref{eq:ent-cond}) reduces to
\begin{equation}
\mathbf g_1^2\mathbf g_2^2\sin^2(\vartheta) \left[\left(\sigma_{22}-|\sigma_{21}|^2\right)^2-N^2|\sigma_{21}|^4\right]<0.
\label{eq.regular}
\end{equation}
%Due to the negative contribution proportional to $N^2$, entanglement can be generated  under rather general conditions. 
It is straightforward to reduce the condition~(\ref{eq.regular})
for the generation of entangled spatial light modes by $N$ atoms, by using the general property $\sigma_{22}{\geq}|\sigma_{21}|^2{\geq}0$, to
\begin{equation}
    \frac{\sigma_{22}}{|\sigma_{21}|^2}< N+1.
\label{Nsq0}
\end{equation} 
Inserting the stationary values of $\sigma_{ij}$ from \cite{Vogel:06}, we rewrite the entanglement condition in terms of the atomic and driving-field parameters and obtain for the Rabi frequency $\Omega_{\rm R}$
\begin{equation}
        \Omega_{\rm R}^2<\left(\frac{N+1}{2}\frac{\Gamma_1^2}{\Gamma_2^2}-
\frac{\Gamma_1}{\Gamma_2}\right)\left(\Gamma_2^2+\Delta^2\right),
\label{Nsq}
\end{equation}
where $\Gamma_1$ and $\Gamma_2$ are the energy and phase relaxation rates, respectively. Together with the deviation $\Delta$ of the laser frequency from the atomic transition, these rates determine the maximal driving field for which entanglement is obtained. 

The Rabi frequency is chosen real, so that the right hand side of the inequality~(\ref{Nsq}) has to be positive.
For $N{=}1$, this condition is equal to the requirement for the possibility to observe squeezing in single atom resonance fluorescence~\cite{walls}. 
For $N{>}1$ the driving field can be increased by a factor of about $\sqrt{N}$ compared with a single atom.
More importantly, with an increasing number of atoms the ratio $\Gamma_2/\Gamma_1$ of the individual atoms can exceed the value of one, being the ultimate limit beyond which squeezing in resonance fluorescence of a regular $N$-atom system disappears~\cite{vowe}. Thus, for such a system entanglement is more robust against dephasing than squeezing.

We would like to emphasize at this point that the resonance fluorescence not only provides a continuously radiating entangled source with high stability against dephasing, but it is also non-Gaussian. This is easily verified, since correlations containing operator products $\hat{\mathbf {E}}^{(\pm)n}_1 \hat{\mathbf {E}}^{(\pm)m}_2$ are zero when $n{+}m$ exceeds the number $N$ of atoms. It is important that non-Gaussianity is ''at times necessary to successfully perform quantum information tasks``, cf.~\cite{walborn} and Refs.~[4-17] therein. 
This includes applications for entanglement distillation and swapping, quantum teleportation, universal quantum computing, and others. That is, for some important quantum tasks the non-Gaussian property is a desired resource. For this reason, protocols have been established for the aim to de-Gaussify intially 
Gaussian entangled radiation fields, see e.g.~\cite{Ourjoumtsev}.
The resonance fluorescence source under study is per se non-Gaussian and hence does not require such additional procedures. This opens a wide field of possible applications whose study is beyond the scope of this Letter.

\begin{figure}[ht]
\includegraphics[width=8cm]{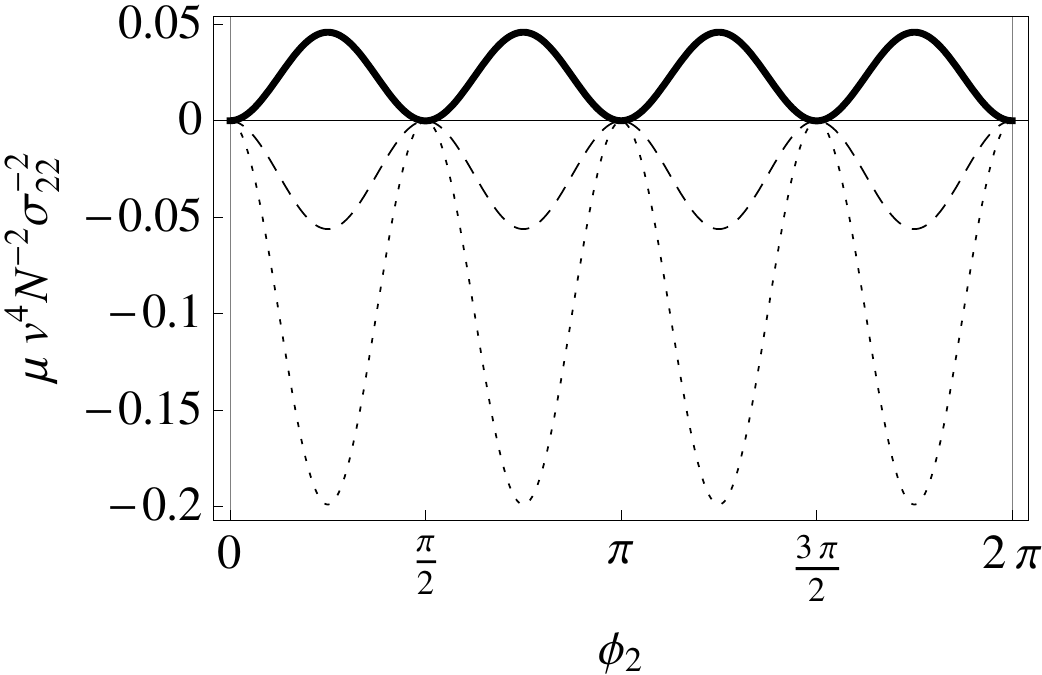}
\caption{The normalized minor $\mu$ is shown, 
as a function of the angle $\phi_2$,
for a linear chain of atoms with $N=2$ (solid), $3$ (dashed), and $4$ (dotted). The pumping strength yields $\sigma_{22}/|\sigma_{21}|^2{=}3.5$.}\label{fig.order}
\end{figure}

In Fig.~\ref{fig.order} we show the minor $\mu$ normalized by $\nu^{-4}N^2\sigma_{22}^2$, cf. Eq.~(\ref{eq.statD1}) together with (\ref{eq.g}), for a linear chain of atoms. All atoms are exactly aligned along the $z$-axis, with an inter-atomic distance of ten wavelengths of the driving laser.
The direction of the laser is chosen to be almost parallel to the $z$-axis, 
so that the Rabi frequencies are kept nearly constant for a sufficiently large number of atoms. The two detectors are positioned in the far field region (at distance  $r$ from the coordinate center), detector 1 in the 
$y$-direction and detector 2 can be moved in the $x$-$y$-plane. Hence, the  position of the latter only depends on the azimuth angle $\phi_2$.
The system parameters are chosen such that the incoherent part of the atomic fluorescence plays a significant role, $\sigma_{22}/|\sigma_{21}|^2{=}3.5$.
We remind the reader that under these conditions squeezing in resonance fluorescence would not occur. According to the condition~(\ref{Nsq0}), entangled light is irradiated by three or more atoms. We consider the situation for the numbers of atoms being  $N{=}2,\ 3, \ 4$. It is clearly seen from Fig.~\ref{fig.order}, that entanglement occurs for $N\ge 3$. 
Even for the chosen relatively poor coherence properties, entanglement of the radiation fields in two directions is found in almost every direction. Note that for opposite directions of the detectors relative to the atomic sample ($\mathbf r_1{=}{-}\mathbf r_2$, i.e. $\sin\vartheta{=}0$) the entanglement vanishes in general. 

The linear chain cannot be experimentally realized to arbitrary precision. For more and more atoms the negativity of the minor increases together with the robustness of the corresponding entanglement against dephasing allowing less accuracy without destroying the entanglement. For the purpose of describing a realistic system, let us consider a linear Paul trap for a chain of ions. Note that present technologies, such as the multizone trap-array architectures designed for quantum computation~\cite{blatt-winel}, may offer alternative possibilities to realize an entangled radiation source of the type under study. The positions of the ions are rather well fixed in the linear trap. However, they are not perfectly equidistant for more than three ions, due to the distance dependence of the Coulomb interaction within the chain of ions, for more details see e.g.~\cite{buzek}. 
The atomic distances are scaled by an externally controlled length parameter $\gamma$, which can be properly adjusted to the number $N$ of trapped ions.

One also cannot ignore fluctuations in the positions, as the individual ions are not really fixed due to the uncertainty principle. Assuming a harmonic potential around each of the equilibrium positions of the ions, the frequency of these harmonic oscillators are nearly equal to the frequency of the trap oscillation. 
This simple assumption overestimates the variances of the atomic positions, since the potential around a given ion rises quickly due to anharmonic effects. Hence the true position uncertainty can be much smaller than that of an ion in the full trap potential. Nevertheless, we choose this larger value, to effectively account for other possible 
experimental imperfections.
Using elementary quantum mechanics and statistics, we can approximate the physical limit of the variance of the individual ions as
\begin{equation}
        \Delta z\approx \sqrt[4]{\frac{4\pi\epsilon_0\hbar^2\gamma^3}{MQ^2}},\label{eq.vartrap}
\end{equation}
where $M$ is the mass of one of the ions and $Q$ its charge. Assuming for  example $Q=e$ and $M=3.3309\times 10^{-25}$ kg and an optical transition wavelength $\lambda=194.2$ nm, we obtain
$\Delta z\approx 1.014\times10^{-9}\mbox{m}\left(\frac{\gamma}{\lambda}\right)^{3/4}$. This is the situation to be realized with mercury ions, cf. the experiments in~\cite{wineland}.
As the variance should be small compared to the transition wavelength to keep statistical averaging effects in the minor $\mu$ given by Eq.~(\ref{eq.statD1}) sufficiently small, we assume a limit of $\Delta z{\leq}0.1\lambda$. This leads to a maximal value for $\gamma$ of about $50 \lambda$. Thus we conclude that linear traps would be suited to provide sufficient positioning accuracy of the atoms to realize a continuously radiating entangled-light source with regularly arranged atoms for average distances large compared to the wavelength.

To verify these theoretical predictions by simulations, we choose the parameter $\gamma$ in such a way, that for each $N$ the minor $\mu$ becomes minimal. The resulting $\Delta z$ from Eq.~(\ref{eq.vartrap}) is implemented as an upper bound for a random deviation from the ideal position along the $z$-axis, variations perpendicular to this axis are assumed to be negligible due to properly chosen potentials.
The minors have been calculated for the same number of ions and the same  parameters as in the ideal case shown in Fig.~\ref{fig.order}. 
The calculations for the realistic chain are in reasonable agreement with the  idealized results given above. For $N=2,3$ the uncertainties of the ion's positions modify the idealized result by less than one percent. For four ions in the linear trap, we could optimize the negativities to about 98.8\% of the ideal case. Thus it is possible to obtain almost optimal entanglement in the atomic resonance fluorescence with ions in a linear trap, as long as the number $N$ of ions is not too large. By applying alternative technologies for realizing regular atomic systems, e.g. by trap arrays~\cite{blatt-winel}, one may also overcome this limitation.

In conclusion, we have studied the realization of a continuous radiation source emitting two multi-mode light beams in different spatial directions, which show bipartite entanglement.
The desired radiation source consists of the atomic resonance fluorescence of a regular system of several non-interacting atoms. This represents a non-Gaussian source of entangled radiation. For a single atom, entanglement is obtained under the same conditions as required for the realization of squeezing in resonance fluorescence.
For a system of $N$ atoms, the situation can be significantly improved, 
as by increasing the number of atoms the entanglement of the emitted light beams is getting more robust against atomic dephasing. As an example, we have estimated the practical limitations of the achievable entanglement for a system of ions in  a linear trap. Altogether, the resonance fluorescence of regular atomic systems is a continuous and non-Gaussian source of entangled light, which may open interesting perspectives for various applications in quantum information technology.

This work was supported by the Deutsche Forschungsgemeinschaft, SFB 652.

\end{document}